\documentclass[a4paper,nofootinbib,11pt]{revtex4}
\usepackage{amssymb,amsmath,amsfonts}
\usepackage{graphicx}
\usepackage{epstopdf}
\usepackage{hyperref}
\begin{document}
\title{Hawking radiation of $E<m$ massive particles in the tunneling formalism}
\author{G. Jannes\footnote{e-mail: jannes@ltl.tkk.fi}}
\affiliation{Low Temperature Laboratory, Aalto University School of Science, PO Box 15100, 00076 Aalto, Finland}
\date{\today}
\begin{abstract}
We use the tunneling formalism to calculate the Hawking radiation of massive particles. For $E \geq m$, we recover the traditional result, identical to the massless case. But \mbox{$E<m$} particles can also tunnel across the horizon in a Hawking process~\cite{Jannes:2011vb}. We study the probability for detecting such $E<m$ particles as a function of the distance from the horizon and the energy of the particle in the tunneling formalism. We derive a general formula  and obtain simple approximations in the near-horizon limit and in the limit of large radii.
\end{abstract}
\maketitle
\section{Introduction.}
Hawking radiation \cite{Hawking:1974sw} consists in the emission of pairs of quanta from a black-hole horizon: one with ``positive energy'' (positive co-moving frequency) towards the exterior and one with ``negative energy'' (negative co-moving frequency) towards the interior of the black hole. The emission has a thermal spectrum as seen by an asymptotic observer at infinity, with a temperature $T_H$ determined by the derivative of the free-fall velocity $v$ at the horizon $r_h$:  $T_H=\hbar\left|\frac{dv}{dr}\right|_{r=r_h}/2\pi$. This result can be obtained very simply in the semiclassical method. This tunneling description of Hawking radiation, first introduced by Volovik \cite{Volovik:1999fc} (see also \cite{Parikh:1999mf} for the Parikh-Wilczek version, \cite{Zhang:2005sf} for its extension to $E>m$ massive particles, and the recent~\cite{Miao:2010xu} for charged $E>m$ massive particles), indicates that the Hawking process can be understood as the quantum tunneling across the horizon between classical trajectories on both sides of the horizon. The semiclassical method was also applied to Hawking radiation from rotating~\cite{Jiang:2005ba}
and charged~\cite{Zhang:2005xt} black holes, and to several higher-dimensional and other more exotic black hole geometries (see e.g. \cite{Matsuno:2011ca,Ren:2010zzc,Mehdipour:2010ap,Dehghani:2010zz}
for some recent examples), as well as to related phenomena such as the Zel'dovich-Starobinsky effect \cite{Zeldovich,Starobinsky} (see \cite{Volovik:2003fe}), and the Unruh effect \cite{Unruh:1976db}
(see e.g.~\cite{deGill:2010nb}). 

We will focus on massive particles in a Schwarzschild geometry. 
If one is interested only in the spectrum as measured at infinity, then it is straightforward to arrive at the usual conclusions~\cite{Page:1976df}, namely that there is a threshold $E\geq m$ for massive quanta, but otherwise, the spectrum does not differ from that of the standard massless case. 

However, although no $E<m$ particles arrive at infinity, they are nevertheless radiated from the horizon, see also~\cite{Kofman:1982gu}. We will calculate the 
probability for detecting such $E<m$ particles as a function of the distance from the horizon and the energy of the particle. We largely follow Volovik's derivation and notation \cite{Volovik:1999fc} (see also \cite{Volovik:2006,Volovik:2009eb}), focus on purely radial movement in a Schwarzschild geometry and set $\hbar=c=1$. 

\section{General calculation.}
We start from the Painlev\'e-Gullstrand-Lema\^itre (PGL) form of the Schwarzschild line element:
\begin{equation}\label{PGL}
ds^2=g_{\mu\nu}dx^\mu dx^\nu =-(1-v^2(r))dt^2- 2v(r)dr dt +dr^2
\end{equation}
where $v(r)=-\sqrt{r_h/r}$ is the free-fall velocity  and $r_h$ the horizon or Schwarzschild radius. The PGL metric is stationary and moreover, unlike the Schwarzschild form, it is regular across the horizon. This makes it particularly suited for the tunneling description of Hawking radiation.

For a massive particle, one has
\begin{equation}
 g^{\mu\nu}p_\mu p_\nu=-m^2
\end{equation}
with $p_\mu=(-E,p_i)$, leading to the energy-momentum dispersion relation 
\begin{equation}\label{disp-rel}
m^2+p^2=(E-\bf p\cdot \bf v)^2.
\end{equation}
Since $v$ represents the free-fall velocity, it is natural to write this as 
\begin{equation}
E=E_0+\bf p\cdot \bf v ,
\end{equation}
and interpret $E$ as the Doppler-shifted energy (the energy in the ``black hole rest frame''), which is a conserved quantity, and $E_0=\sqrt{m^2+p^2}$ as the energy in a co-moving reference frame. This interpretation is reinforced by the observation that the PGL metric is also the natural metric in analogue gravity~\cite{Barcelo:2005fc}, where it describes the propagation of sound waves or other perturbations on a background fluid moving with a velocity $v$.

We want to calculate the influence of the mass on the tunneling rate of particles from the black hole up to a detector (where the particle can be captured in a bound state, see also~\cite{Castineiras:2002zf}) at an arbitrary fixed radial distance $R$. The tunneling probability $W$ is determined by the imaginary part of the action $S$ along the semiclassical trajectory: 
\begin{equation}
W\propto \exp[-2\text{Im}S],
\end{equation}
where
\begin{equation}
\text{Im}S=\text{Im}\int p_r(r)dr 
\end{equation}
and $p_r(r)$ follows from the dispersion relation \eqref{disp-rel}. The final result will be of the form 
\begin{equation}\label{expected-result}
W(E)\propto \exp[-2\text{Im}S_1]\exp[-2\text{Im}S_2] 
\end{equation}
with $\text{Im}S_1$ a contribution for tunneling through the horizon, and $\text{Im}S_2$ an additional action in case there is a second classically prohibited region beyond the horizon. We certainly expect $\text{Im}S_2\neq 0$ when $E<m$, since the massive particle is then classically forbidden [see \eqref{disp-rel}] in flat spacetime [$v(r \to \infty)=0$].

From~\eqref{disp-rel}, we obtain
\begin{align}
p
&=-\frac{Ev}{1-v^2} + \frac{1}{1-v^2}\sqrt{m^2(v^2-1)+E^2}\\
&=p_1+p_2
\end{align}

Note that, in order for the semiclassical formalism to be valid,  the action must be large ($\text{Im}S\gg 1$). In particular, this implies that we consider the case $m r_h\gg 1$. In the opposite limit $m r_h\ll 1$, it was found in~\cite{Kofman:1982gu} that the created bound particles are localized mainly on resonant levels outside the black hole with large occupation numbers.

\subsection{$S_1$: Tunneling through the horizon.} 
For $p_1$, we shift the contour of integration to the complex plane and apply the standard residue theorem $\int_C f(z)=2\pi i\text{Res} f$. There is one pole along the radial path: $v=-1$ at $r=r_h$, so
\begin{equation}
\text{Res} p_1=\text{Res}\frac{-Ev}{(1+v)(1-v)}=\frac{E}{2v'(r_h)},
\end{equation}
where the prime denotes $d/dr$. 
We recover the standard Hawking result, as for a massless particle to reach infinity: \mbox{$2\text{Im}S_1=E/T_H$, with $T_H=|v'(r_h)|/2\pi$.} The presence of a mass has absolutely no influence on the probability of tunneling across the horizon.

\subsection{$S_2$: Tunneling towards the detector at a distance $R$.}
$p_2$ will only have an imaginary contribution when
\begin{equation}\label{condition}
E^2<m^2(1-v^2).
\end{equation}
For a Schwarzschild black hole ($v^2=r_h/r$), this leads to
\begin{equation}\label{E_c}
E<E_c(R)=m\left(1-\frac{r_h}{R} \right)^{1/2}
\end{equation}
for given $R$. Alternatively, it
can be expressed as a condition on $R$ for given $E$:
\begin{equation}\label{rc}
R>r_c(E)=\frac{m^2}{m^2-E^2}r_h= \frac{1}{1-\frac{E^2}{m^2}}r_h
\end{equation}
For $E>E_c$ (or $R<r_c$), there is no additional barrier. Once the particle has tunnelled through the horizon, it can freely propagate up to the detector at $R$. At sufficient distance from the black hole ($v\to 0$), the condition for a second imaginary contribution reduces to $E<m$, as we anticipated. This means that, for $E\geq m$, the mass term does not cause any additional tunneling factor, independently of $R$, and we recover the standard result for massless particles all the way to $r \to \infty$. For $E<m$, however, the particle is created with an energy $E$ which is insufficient to escape all the way to $\infty$, so it will encounter a second barrier as $v$ decreases (i.e., as the particle moves towards flat spacetime).

In case of \eqref{condition}, we take the (positive) imaginary part and obtain 
\begin{align}
\text{Im}S_2
&=\int_{r_c}^R dr \frac{\sqrt{m^2-E^2}}{1-v^2}\sqrt{1- v^2\frac{m^2}{m^2-E^2}}\label{integral-2}\\
&=\int_{r_c}^R dr \frac{1}{1-r_h/r}\sqrt{m^2(1-r_h/r)-E^2}\label{integral-Schw}
\end{align}
where the last expression is specific for a Schwarzschild profile.

Note that the logarithmic divergence for $r\to r_h$ ($v\to 1$) is avoided because $r_c>r_h$. 

\begin{figure}
\includegraphics[width=0.45\textwidth]{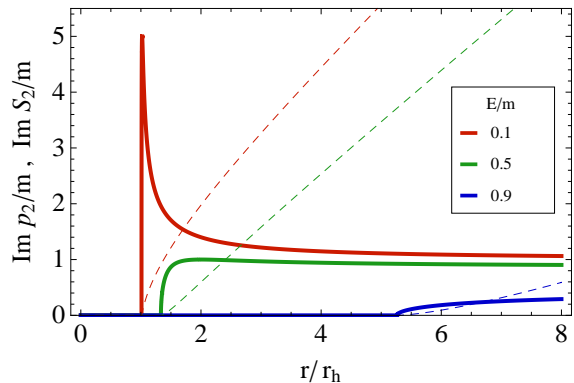}
\caption{\label{fig:potential} 
Fig.\ref{fig:potential}: Semiclassical barrier Im$p_2$ (thick lines) and tunneling action $\text{Im}S_2$ (dashed lines) beyond the horizon $r_h$ for various values of $E/m<1$.
}
\end{figure}

Im$p_2$ is shown, together with Im$S_2$, for various values of $E/m$ in 
Fig.\ref{fig:potential}. For $E/m<1/\sqrt{2}$, Im$p_2$ increases to a maximum $\frac{1}{2}m^2/E$ at $r=r_hm^2/(m^2-2E^2)$ before decreasing to the asymptotic value $\sqrt{m^2-E^2}$ for $r\to \infty$. For $E/m\geq 1/\sqrt{2}$, the maximum disappears and the increase is monotonic.
%
%

\section{Limit cases}
Simple analytic results can be found in the following interesting limit cases.
\subsection{Limit $R\gg r_h$}
\begin{figure}
\includegraphics[width=0.45\textwidth]{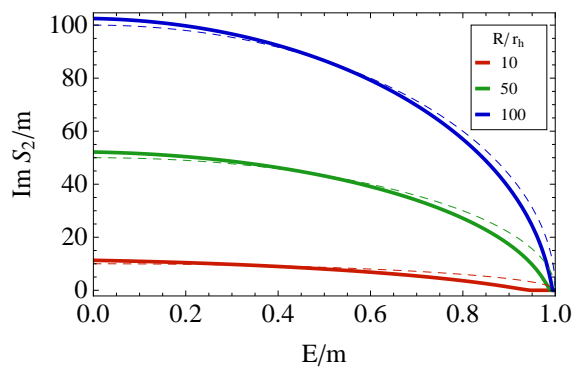}
\caption{\label{fig:limit-large-R} 
Fig.\ref{fig:limit-large-R}: Tunneling action $\text{Im}S_2$ (thick lines) and approximation \eqref{large-R} for $R\gg r_h$ (dashed lines) as a function of $E/m$.
}
\end{figure}

When the detector is very far from the horizon, the integral is dominated by the contributions where $v\to 0$ and we can immediately write 
\begin{align}\label{large-R}
\text{Im}S_2=\sqrt{m^2-E^2}\int_{r_c}^R dr\approx R \sqrt{m^2-E^2}
\end{align}
where we have assumed $R\gg r_c$, see Fig.\ref{fig:limit-large-R}.
The global tunneling probability 
\begin{equation}\label{result-large-R}
W(E)\propto \exp\left[\frac{-E}{T_H}\right]\exp\left[-2R\sqrt{m^2-E^2}\right]
\end{equation}
decreases exponentially with the distance $R$ to the detector, and the energy difference $\sqrt{m^2-E^2}$.
%
%
Note that this result does not depend on the Schwarzschild profile, but is generally valid as long as $R$ lies sufficiently far away from the horizon for the integral~\eqref{integral-2} to be dominated by the region $v\ll 1$. 


\subsection{Limit $R\to r_h$}
Another interesting limit is $v\to 1$, i.e. $\frac{R}{r_h}-1\ll 1$. 
In this near-horizon limit, there is a non-zero contribution only for $E\ll m$, see \eqref{rc}. There will then be a strong barrier almost as soon as the particle crosses the horizon (see the case  $E/m=0.1$ in Fig.\ref{fig:potential} above), since it has practically no energy to sustain its own mass. 

Starting from \eqref{integral-Schw}, we write $r/r_h-1=\frac{E^2}{m^2}y\ll 1$, and obtain
\begin{equation}
 \text{Im}S_2=Er_h\int_{1}^{\frac{m^2}{E^2}(R/r_h-1)} \frac{dy}{y} \sqrt{y-1}
\end{equation}
where we have used $\frac{E^2}{m^2}\ll 1$ and $r\approx r_h$.

Two limit cases yield a simple interesting result.
\textbullet~ $\frac{E^2}{m^2}\ll\frac{R}{r_h}-1\ll 1$ (i.e. $\ \frac{R}{r_c}-1\gg \frac{E^2}{m^2}$)

\begin{figure}
\includegraphics[width=0.45\textwidth]{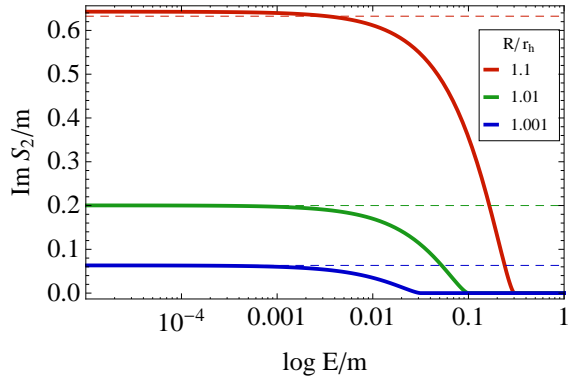}
\caption{\label{fig:limit-small-r-case1} 
Fig.\ref{fig:limit-small-r-case1}: Tunneling action $\text{Im}S_2$ (thick lines) and approximation \eqref{approx-small-R-case1} for $E\ll E_c$ (dashed lines) as a function of $E/m$ for different positions $R$ of the detector near the horizon.
}
\end{figure}

For given $R$, this implies $E^2/E_c^2(R)\ll 1$. 
The integral is then dominated by the region where $y\gg 1$ so 
\begin{align}\label{approx-small-R-case1}
 \text{Im}S_2&\approx
2mr_h\left(\frac{R}{r_h}-1\right)^{1/2}
\end{align}
The second tunneling action $\text{Im} S_2$ becomes independent of the energy $E$ in this limit, as illustrated in Fig.\ref{fig:limit-small-r-case1}.
%
%

\textbullet~ $0<\frac{m^2}{E^2}\left(\frac{R}{r_h}-1\right)-1\ll 1$, (i.e. $\ \frac{R}{r_c}-1\sim \frac{E^2}{m^2}$)


For given $R$, this condition is equivalent to \mbox{$1-\frac{E^2}{E_c^2}\ll 1$}. 
Now $y\approx 1$ in the whole integration region, so
\begin{align}
 \text{Im}S_2&\approx
\frac{2}{3}r_h\frac{m^3}{E^2}\left[\left(\frac{R}{r_h}-1\right)-\frac{E^2}{m^2} \right]^{3/2}\\
&\approx\frac{2}{3}r_hE_c\left(1-\frac{E^2}{E_c^2}\right)^{3/2}\\
&\approx\frac{2}{3}mr_h\left(\frac{R}{r_h}-1\right)^{1/2}\left(1-\frac{E^2}{E_c^2}\right)^{3/2}\label{approx-small-R-case2}
\end{align}
where we have used $R\approx r_h$.

\begin{figure}
\includegraphics[width=0.45\textwidth]{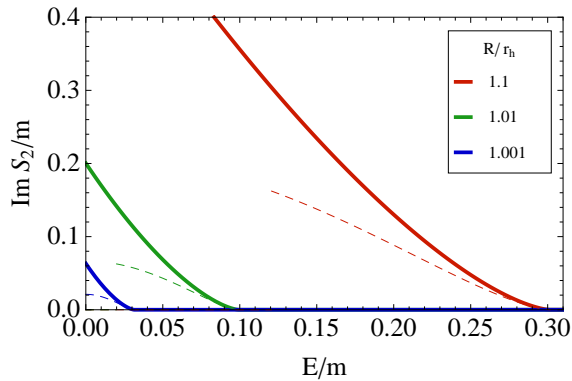}
\caption{\label{fig:limit-small-r-case2} 
Fig.\ref{fig:limit-small-r-case2}: Tunneling action $\text{Im}S_2$ (thick lines) and approximation \eqref{approx-small-R-case2} for $E\sim E_c$ (dashed lines) as a function of $E/m$ for different positions $R$ of the detector near the horizon.
}
\end{figure}

The prefactor is similar to the previous result, but now there is an additional suppression with decreasing $E$, see Fig.\ref{fig:limit-small-r-case2}.
%

\section{Conclusion.}
Massive particles tunnel across the horizon at exactly the same rate as massless particles of the same energy $E$, even if $E<m$. Such massive particles with $E<m$ do not reach infinity, but nevertheless have a non-zero probability of being detected at any finite distance $(R-r_h)$ from the horizon. For a detector very close to the horizon, or very far from the horizon, the detection probability reduces to simple expressions in terms of $E/m$ and $R/r_h$. Different laboratory systems in which this behaviour of massive particles near a black hole could be simulated in the context of analogue gravity were suggested in \cite{Jannes:2011vb}. An interesting prospect for future work is that the double barrier could cause a resonance mechanism, thereby leading to peaks of strongly increased tunneling probability. Such resonant Hawking radiation in the presence of double-barrier structures was recently studied for a BEC-based analogue system in~\cite{Zapata:2011ze}.

\section*{Acknowledgements.}
It is a pleasure to thank G.~E. Volovik for encouragement, clarifying discussions and help on the calculations. The author is supported by a FECYT postdoctoral contract of the Spanish Ministry of Education. This work is also supported in part by the Academy of Finland and its CoE program 2006--2011

\end{document}